\def\ba{\begin{eqnarray}}

\def\ea{\end{eqnarray}}

\def\l{\label}

\def\n{\nonumber \\}

\def\b{\bibitem}

\documentclass[12pt]{article}

\begin{document}

\title{Coincidence probability as a measure
of the average phase-space density at freeze-out}

\author{A.Bialas, W.Czyz and K. Zalewski \\ M.Smoluchowski
Institute of Physics \\Jagellonian University,
Cracow\thanks{Address: Reymonta 4, 30-059 Krakow, Poland;
e-mail:bialas@th.if.uj.edu.pl;}
\\ Institute of Nuclear Physics, Polish Academy of Sciences
\thanks{Radzikowskiego 152, Krakow, Poland}}

\maketitle

\begin{abstract}

It is pointed out that the average semi-inclusive particle
phase-space density at freeze-out can be determined from the
coincidence probability of the events observed in multiparticle
production. The method of measurement is described and its
accuracy examined.

\end{abstract}

{\bf 1.} Recently, several methods were proposed which allow to
estimate the average of the single-particle {\it inclusive}
phase-space density produced in ultrarelativistic heavy ion
collisions \cite{ber1}-\cite{sk}. This quantity is useful in
discussions of the equilibrated systems and therefore such
measurements open possibilities to verify the expected presence of
the thermal phase-space distribution at freeze-out \cite{heinz1}
and/or search for more exotic phenomena \cite{led}.

In the present paper we propose an extension of these studies by
including, in addition, the {\it exclusive} and {\it
semi-inclusive} M-particle phase-space densities. We show that
their averages can be estimated from measured coincidence
probabilities of the multiparticle events observed in high-energy
collisions. The information one may gain from this approach is
complementary to that obtained from single-particle inclusive
measurements. In particular, it gives an insight into the
correlation structure of the final state of the collision
(including both dynamic and Bose-Einstein correlations), a feature
which is ignored in the single-particle inclusive measurements
described in \cite{ber1}-\cite{sk}. Furthermore, one can show
\cite{bcwe} that the average semi-inclusive phase-space densities
are closely related to the second Renyi entropy \cite{ren} and
thus their measurement allows to estimate a lower limit of the
true (Shannon) entropy of the system (without assuming the
thermodynamic equilibrium). Needless to say, a comparison of the
measured average semi-inclusive phase-space densities with
expectations for the thermalized systems would be very interesting
indeed.

As emphasized in \cite{ber1,bpb}, the phase-space distribution of
particles produced in high-energy collisions is not a precisely
defined quantity. Apart from the standard problems with the
uncertainty principle, one has to take into account that particles
may be produced at different times \cite{ber1}. Following Bertsch
\cite{ber1}, we shall ignore this problem and assume that all
particles are created at the same time. That is to say, we are
considering a time-average of the density \cite{ber1,bpb}.

To define the average semi-inclusive phase-space density, consider
a collection of events in which exactly $M$ particles were
observed in a given region of the momentum space. We shall call
them $M$-particle events (independently of how many particles were
actually produced)\footnote{This is terminology often used in
experimental description of multiparticle processes. The proper
technical term is the exclusive distribution if all particles are
observed, and semi-inclusive distribution if besides a given
number of observed particles there is an unspecified number of
other particles. This should not be confused with inclusive
$M$-particle distributions.}. These events can be described by the
normalized $M$ particle phase-space distribution $W_M(X,K)$, with
$X=X_1,...,Z_M$, $K=K^{(1)}_x,...,K^{(M)}_z$. The corresponding
particle phase-space density is $D_M(X,K)=MW_M(X,K)$ and thus

\ba <| D_M|> = M \int dX dK |W_M(X,K)|^2 \equiv M(2\pi)^{-3M} C_M.
\l{7c} \ea gives the average phase-space density of the $M$
particle system.

It should be emphasized that, as is clear from this discussion,
the phase-space density $D_M$, describing the semi-exclusive
distribution, refers only to particles actually {\it measured} in
a given experiment and in a given mometum region. It gives no
direct information about the particles which are not registered in
the detector. To obtain information on the phase-space density of
{\it all} produced particles additional assumptions (e.g.
thermodynamic equilibrium) are necessary.

Note that the average semi-inclusive phase-space density averaged over
 all particle multiplicities is simply obtained from (\ref{7c}), using
\ba <<D>>=\sum_M P(M) <D_M>       \l{77c} \ea where $P(M)$ is the
multiplicity distribution. Therefore from now on, to simplify the
discussion, we shall only consider the case of a fixed
multiplicity.

Our method is based on the observation that, for a rather wide
class of models of particle production, the quantitity $C$,
defined in (\ref{7c})\footnote{To simplify the formulae we shall
from now on omit the index $M$ in all quantities. Since we are
discussing solely $M$-particle events, this should not lead to any
confusion.}, can be approximated by the measured coincidence
probability $C^{exp}$ of the events with $M$ particles, defined as
\cite{bcw,bc1} \ba C^{exp} =\frac {N_2}{N(N-1)/2} \l{31} \ea where
$N_2$ is the number of the observed pairs of  identical events and
$N$ is the total number of events. $N(N-1)/2$ is the total number
of pairs of events\footnote{Formula (\ref{31}) was first
suggested, in a different context, by Ma \cite{ma}.}.

It is clear that, since the observed events are described by the
particle momenta which are continuous variables, Eq. (\ref{31}) is
not directly applicable: a discretization is necessary. Then one
can define the identical events as those which have the same
population of the predefined bins and thus counting of
coincidences becomes straightforward\footnote{A detailed
description of this procedure was given in \cite{dl} and applied
in \cite{kit}.}. The counting of identical events obviously
depends on the binning, so the procedure is ambiguous
\cite{bcw,bc1,dl,bc2}. In order to obtain a viable estimate of the
average particle density, we thus have to select the binning in
such a way that the result of (\ref{31}) is as close as possible
to the exact value of $C$ which, as seen from (\ref{7c}), gives
directly the particle phase-space density.

In the present paper we argue that for a fairly large class of
physically sensible models, one can find an adequate binning
procedure and thus to determine rather precisely $<D>$ by
measuring $C^{exp}$, i.e., by counting the number of pairs of
identical events. The method turns out to be particularly suitable
for large systems and thus may be useful in heavy ion collisions.

In the next section the discretization procedure is described in
some detail and the corresponding formulae for $C^{exp}$ are
written down. The phase-space density $<D>$ and its relation to
$C^{exp}$ are discussed in Section 3. Our conclusions and outlook
are given in the last section.

{\bf 2.} In this section we discuss how the discretization
procedure afects the definition (\ref{31}) of the coincidence
probability. To this end we first express the $C^{exp}$ given by
(\ref{31}) in terms of the momentum distribution of $M$ particles
$w(K)=w(K^{(1)},...,K^{(M)})$.

Consider a set of discretized events constructed by dividing the
particle momentum space into $J$ rectangular bins of volume

\ba \omega_j=(\Delta_x\Delta_y\Delta_z)_j\;;\;\;j=1,...,J. \l{33a}
\ea Then the probability to find a particle in  bin
$\omega_{j_1}$, another one in  bin $\omega_{j_2}$, etc. is \ba
P(j_1,j_2,...,j_M)= \prod_{m=1}^M\omega_{j_m}
<w\left(K_{j_1}^{(1)},...,K_{j_M}^{(M)}\right)> \l{s1} \ea where
\ba <w\left(K_{j_1}^{(1)},...,K_{j_M}^{(M)}\right)>=\n=
\prod_{m=1}^M(\omega_{j_m})^{-1}
\int_{\omega_{j_1}}dK^{(1)}...\int_{\omega_{j_M}}dK^{(M)}
w(K^{(1)},...,K^{(M)}).  \l{s2} \ea Note that the bins
$\omega_{j_1},...,\omega_{j_M}$ do not have to be different.

Thus the coincidence probability as measured by the formula
(\ref{31}) is

\ba C^{exp}_M=\sum_{j_1}...\sum_{j_M}
\left[P(j_1,j_2,...,j_M)\right]^2=\n=
\sum_{j_1}...\sum_{j_M}\prod_{m=1}^M[\omega_{j_m}]^2
[<w\left(K_{j_1}^{(1)},...,K_{j_M}^{M}\right)>]^2    \l{s3} \ea
The first equality follows from the observation that sampling a
series of  events is the Bernoulli process and thus probability to
find, after $N$ trials,
 $n_1,...,n_J$ events in configurations $\{1\},...\{J\}$  is
\ba B(n_1,...n_J)=\frac{N!}{n_1!...n_J!} (P_1)^{n_1}...(P_J)^{n_J}
\l{32} \ea From this formula it is not difficult to see that \ba
<N_2>\equiv \sum_{n_1,...,n_J}\sum_{j=1}^J \frac{n_j(n_j-1)}{2}
B(n_1,...n_J) = \frac{N(N-1)}2 \sum_{j=1}^J (P_j)^2 \l{33} \ea

The question now is: how to select the bins $\omega_j$ to obtain a
result as close as possible to $C$ giving the average value of the
particle phase-space density $<D>$ [c.f. (\ref{7c})]. This is
discussed in the next section.

{\bf 3.} To analyze  the relation between $C^{exp}$ and $C$ we
consider
 the $M$-particle phase-space distribution of the general form
\ba W(X,K)= \frac1{(L_xL_yL_z)^M}G[X/L] w(K)           \l{36a} \ea
with $X/L\equiv (X_1-\bar{X}_1)/L_x,...,(Z_M-\bar{Z}_M)/L_z$,
 $K= K_1,...,K_{M}$. The function $G$
satisfies  the normalization conditions

\ba \int d^{3M}u G(u)&=& 1\;\;\rightarrow\;\;\int dX G(X/L) =
(L_xL_yL_z)^M ;\n \int d^{3M}u u_i G(u)&=& 0\;\;\rightarrow\;\;
<X_i,Y_i,Z_i>= \bar{X}_i,\bar{Y}_i,\bar{Z}_i;\n \int
d^{3M}u(u_i)^2 G(u)&=& 1\;\;\rightarrow\;\;
<(X_i-\bar{X}_i)^2,...> =L_x^2,...   \l{37} \ea

The first condition insures that $w(K)$ is the observed
(multidimensional) momentum distribution\footnote{By definition,
$w(K)= \int dX W(X,K)$.}, the second defines the central values of
the particle distribution in configuration space and the third
defines $L_x,L_y,L_z$ as giving root mean square sizes of the
distribution in configuration space. Both sizes and central
positions may depend on the particle momenta\footnote{They may be
also different for different kinds of particles.}. The form of the
function $G$ describes the shape of the multiparticle distribution
in configuration space.

Ansatz (\ref{36a}) for the time-averaged phase space density is
satisfied in a large variety of models \cite{mod}.

Using (\ref{36a}) we obtain from (\ref{7c})

\ba C=(2\pi)^{3M} \int d^{3M}K [w(K_1,...,K_M)]^2 \int\frac{
dX_1...dZ_M} {(L_xL_yL_z)^{2M}} [G(X/L)]^2=\n= (2\pi g)^{3M} \int
d^{3M}K [w(K_1,...,K_M)]^2  \frac1{(L_xL_yL_z)^M} \l{38} \ea with
\ba g^{3M}=\int d^{3M}u [G(u)]^2      \l{39} \ea

The constant $g$ depends on the shape of particle distribution in
configuration space. This dependence is, however,  rather mild.
For example, we obtain $g^{-1}=2\sqrt{\pi}$ for Gaussians and
$g^{-1}=2\sqrt{3}$ for a rectangular box.

In the discretized form,  (\ref{38}) can be written as

\ba C= (2\pi g)^{3M} \sum_{j_1,...,j_M}
\prod_{m=1}^M\frac{\omega_{j_m}}{(L_xL_yL_z)_{j_m}}
<[w_{j_1,...,j_M}]^2> \l{40} \ea where

\ba <[w_{j_1,...,j_M}]^2>=\n= \prod_{m=1}^M(\omega_{j_m})^{-1}
\int_{\omega_{j_1}}dK^{(1)}...\int_{\omega_{j_M}}dK^{(M)}
\left\{w(K^{(1)},...,K^{(M)})\right\}^2  \l{36} \ea and
$(L_xL_yL_z)_{j_m}$ is a suitable average of $(L_xL_yL_z)$ over
 bin $j_m$.

Comparing (\ref{40}) with (\ref{s3})  one sees that to have
$C^{exp}_M$ as close as possible to $C$, the best volume of the
bins is \ba \omega_{j_m}=(\Delta_x\Delta_y\Delta_z)_{j_m}=
\frac{(2\pi g)^3} {(L_xL_yL_z)_{j_m}} \l{41} \ea

One sees from this formula that $\omega$ depends crucially on the
volume of the system in configuration space. One sees,
furthermore, that with this choice of $\omega$ the coincidence
probability $C^{exp}$, determined by counting the number of pairs
of identical  events [c.f. (\ref{31})]$^4$, is related to $C$,
giving directly the average particle phase-space density $<D>$
[c.f. (\ref{7c}) and (\ref{40})], by the formula

\ba C= C^{exp} \frac{\sum_{j_1,...,j_M}<[w_{j_1,...,j_M}]^2> }
{\sum_{j_1,...,j_M}[<w_{j_1,...,j_M}>]^2} \l{42} \ea

It is thus clear that the accuracy of determination of $C$
increases with increasing volume of the system. Indeed, for a
volume large enough, the bins defined by (\ref{41}) are small and
the ratio on the R.H.S. of (\ref{42}) approaches unity. For
smaller volumes the method is less accurate but one may try to
estimate the correcting ratio from the (measured) single particle
distribution.

{\bf 4.} Several comments are in order.

(i) One sees from (\ref{41}) that the optimal size of the bin does
not depend on the average position of the particles at freezeout.
This implies that the momentum-position correlations induced by
the $K$-dependence of $\bar{X}$ do not influence significantly the
measurement of the coincidence probability.

(ii) It is also seen from (\ref{41}) that only the volume of the
bin $\omega_{j_m}=(\Delta_x\Delta_y\Delta_z)_{j_m}$, but not its
shape, matters in the determination of the optimal discretization.
One can use this freedom to improve the accuracy of the
measurement by taking bins large in the directions with weak
momentum dependence and small in the direction where the momentum
dependence is significant.

(iii) One may improve the accuracy of the measurement by
estimating the ratio $\{\sum_{j_1,...,j_M}<[w_{j_1,...,j_M}]^2>
\}/ \{\sum_{j_1,...,j_M}[<w_{j_1,...,j_M}>]^2\} $. This may be
possible if the momentum distribution of particles is measured
with good accuracy.

(iv) Our analysis can be applied to any part of the momentum
space. This allows to measure the local particle density in
momentum space, averaged over all configuration space. In case of
strong momentum-position correlations, the selection of a given
momentum region can induce, however, a selection of a
corresponding region in configuration space.

(v) The accuracy of the measurement depends crucially on the
correct estimate of the size of the system. Information from HBT
measurements should allow to determine the parameters
$L_x,L_y,L_z$ and -at least in principle- also the
shape\footnote{Needed to evaluate the parameter $g$ [cf.
(\ref{39})]. Fortunately, as we already noted, the sensitivity of
$g$ to the shape of $G(u)$ is rather mild.} of the function $G(u)$
(some procedures are described in \cite{dan, sa}). Therefore good
HBT data are essential for a successful application of the method.

(vi) The presented analysis of the discretization procedure can be
generalized to higher order coincidence probabilities \cite{bck}.
This opens the way to a determination of higher Renyi entropies
\cite{ren} and then, by extrapolation, to obtain information on
the Shannon entropy of the system \cite{bc1}.

In conclusion, we propose to estimate the phase-space density of
particles produced in high-energy collisions by measuring the
coincidence probability of the observed events. The accuracy of
the determination of the coincidence probability by counting the
number of the identical events \cite{bcw,bc1} was analysed for a
large class of physically sensible models. It was shown that the
accuracy improves with increasing volume of the system and,
therefore, the method is particularly suitable for heavy ion
collisions. A formula giving the optimal discretization method in
terms of the size of the system in the configuration space [Eq.
(\ref{41})] was derived.

\vspace{0.3cm}

{\bf Acknowledgements}

Discussions with Robi Peschanski and Jacek Wosiek are highly
appreciated. This research was supported by the MNI Grant No 1P03B
045 29.

\end{document}